\documentstyle[psfig,emulateapj]{article}

\def\boldr{\mbox{\boldmath $r$}}

\def\etal{et al.\,}

\def\refnew#1{(\ref{#1})}
\def\be{\begin{equation}}
\def\ee{\end{equation}}

\def\s{\, \rm s}
\def\km{\, \rm km}

\def\g{\rm g}
\def\pomega{\varpi}
\def\deg{^\circ}

\begin{document} 
	
\title{\mbox{Tidal Evolution Of the Planetary System around HD
$83443$}}


\author{Yanqin Wu\altaffilmark{1}, Peter Goldreich\altaffilmark{2}}

\altaffiltext{1}{Canadian Institute of Theoretical Astrophysics,
	University of Toronto, 60 St. George Street, Toronto, ON
	M5S 3H8, Canada, wu@cita.utoronto.ca}
\altaffiltext{2}{Division of Geophysics and Planetary Sciences, 
Caltech, Pasadena, CA 91125, USA, pmg@gps.caltech.edu}

\begin{abstract}
Two planets with orbital period ratio approximately 10:1 have been
discovered around the star HD $83443$. The inner and more massive
planet, HD $83443b$, has the smallest semi-major axis among all
currently known exoplanets. Unlike other short period exoplanets, it
maintains a substantial orbital eccentricity, $e_1=0.079\pm 0.008$, in
spite of efficient tidal damping. This is a consequence of its secular
interactions with HD $83443c$ whose orbital eccentricity $e_2=0.42\pm
0.06$. Dissipation, associated with tides the star raises in the inner
planet, removes energy but not angular momentum from its orbit, while
secular interactions transfer angular momentum but not energy from the
inner to the outer planet's orbit.  The outward transfer of angular
momentum decreases the tidal decay rate of the inner planet's orbital
eccentricity while increasing that of the outer planet.  The alignment
of the apsides of the planets' orbits is another consequence of tidal
and secular interactions. In this state the ratio of their orbital
eccentricities, $e_1/e_2$, depends upon the secular perturbations the
planets exert on each other and on additional perturbations that
enhance the inner planet's precession rate. Tidal and rotational
distortions of the inner planet along with general relativity provide
the most important of these extra precessional perturbations, each of
which acts to reduce $e_1/e_2$. Provided the planets' orbits are
coplanar, the observed eccentricity ratio uniquely relates $\sin i$
and $C\equiv (k_{2}/k_{2J})(R_1/R_J)^5$, where the tidal Love number,
$k_2$, and radius, $R_1$, of the inner planet are scaled by their
Jovian equivalents.
 
\end{abstract}


\setcounter{equation}{0}

\section{Introduction}
\label{sec:introduce}

\begin{deluxetable}{crrrcc} 
\tablecolumns{6} 
\tablewidth{0pt} 
\tablecaption{Parameters for the two planets of HD $83443$.\tablenotemark{a}
\label{table:parameter}}
\tablehead{ 
\colhead{Planet}    & \colhead{P (d)} &   \colhead{$e$}   & 
\colhead{$\pomega (\deg)$} & \colhead{$a$ (AU)} & 
\colhead{$m\sin i$}}
\startdata 
HD$83443b$ & $2.9853\pm 0.0009$ & $0.079\pm 0.008
$ & $300\pm 17$ & $0.0376$  & $1.14$ \\
HD$83443c$ & $29.83\pm 0.18$ & $0.42\pm 0.06$ &
$337\pm 10 $ &  $0.175$  &
$0.53$ \\
 \enddata
\tablenotetext{a}{Data taken from Mayor \etal
(\cite{mayor83443}). Planet masses are in unit of Saturn mass $M_{\rm
Sat} \approx 5.68\times 10^{29}\g$. The host star $M_* = 0.79
M_\odot$. }
\end{deluxetable}

Parameters of the HD $83443$ planetary system as given by Mayor \etal
(\cite{mayor83443}) are listed in Table \ref{table:parameter}.  The
substantial eccentricity of the inner planet's orbit, $e_1 = 0.079 \pm
0.008$, is surprising since other exoplanets with comparable
semi-major axes have circular orbits.  In the absence of other
interactions, tidal dissipation would erode $e_1$ as $d\ln e_1/dt = -
1/\tau$ with (Goldreich \cite{goldreich}, Hut \cite{hut}),
\begin{equation}
\tau \approx {2\over{27}}{{ Q}\over{k_{2}}} \left({{a_1^3}\over{G
M_*}}\right)^{1/2} \left({{m_1}\over{M_*}}\right)
\left({{a_1}\over{R_1}}\right)^5.
\label{eq:dampingtau}
\end{equation}
Adopting values appropriate for Jupiter: radius $R_1\approx 7\times
10^4\km$, mass $m_1\approx 2\times 10^{30}\g$, tidal Love number
$k_{2} \approx 0.5$ (Yoder \cite{yoder}), and tidal quality factor, $Q
\sim 3 \times 10^5$ (Goldreich \& Soter \cite{goldreichsoter}), yields
$\tau \sim 3\times 10^8$ yr.  This is in striking contrast to the old
age of the star. Isochrone fitting suggests an age of about $6.5\,$Gyr
(N. Murray, private communication), which is consistent with the
absence of chromospheric activity and the small projected rotational
velocity, $v \sin i < 2 \km/\s$.

Secular interactions with HD $83443c$ are undoubtedly responsible for
the persistence of HD $83443b$'s orbital eccentricity (Mayor \etal
\cite{mayor83443}). Results from calculations by Lasker and
Benz displayed in Mayor \etal (\cite{mayor83443}) show that secular
interactions between the inner and outer planets induce anti-correlated
oscillations in the two planets' orbital eccentricities. The
anti-correlation results because secular interactions transfer angular
momentum, $J$, but not energy, $E$, and $e$ is
related to $E$ and $J$ by
\begin{equation}
e^2=1+{2J^2E\over m^3(GM)^2}.
\label{eq:eJE}
\end{equation}
Similar phenomena are noted in the hierarchical triple star
systems HD $109648$ (Jha \etal \cite{jha}), HD $284163$ (Griffin \&
Gunn \cite{griffin}), and Algol (see, e.g., Kiseleva \etal
\cite{kiseleva}).

We argue that eccentricity oscillations were tidally damped a long
time ago, and that currently the planets maintain apse alignment (\S
\ref{sec:linear}). In this state the orbital eccentricity of the inner
planet is proportional to that of the outer planet, and both are
steady on the secular time-scale. On the much longer tidal time-scale,
the individual eccentricities decay, with the decay rate being faster
for the inner planet.

We utilize the properties of this system to constrain parameters that
are not directly measurable. From the ratio of the two
eccentricities, we place bounds on the inclination angle, $i$, of the
planets' orbit plane and on the radius of the inner planet.

\section{Consequences Of Apse Alignment}
\label{sec:nonlinear}

\subsection{Assumptions}

We study secular interactions in $HD 83443$'s planetary system under
the assumptions that the planets share a common orbital plane and that
their apsides remain aligned. The former assumption seems reasonable
for planets formed in a gaseous disk. The latter assumption is
consistent with the currently observed approximate apse alignment (see
Table \ref{table:parameter}) and is justified by the analysis in \S
\ref{sec:linear}. These assumptions allow us to infer parameters of
the planetary system from the observed eccentricity ratio.

\subsection{Secular Dynamics}
\label{subsec:currenteratio}

Mutual secular perturbations of the planets' orbits are calculated
following a method devised by Gauss (\cite{gauss}) as outlined in the
appendix. Equation \refnew{eq:gauss} describes the resulting periapse
and eccentricity variations. However, as a result of inner planet's
proximity to the star, it is subject to additional precessional
perturbations of which the most important arise from its related tidal
and rotational distortions (Sterne \cite{sterne})\footnote{The symbol
$k_2$ also appears as the apsidal motion constant in stellar
literature in which case it is a factor 2 smaller than the Love
number.}, and from general relativity (Einstein
\cite{einstein}). Expressions for these precession rates are, 
\begin{mathletters}
\begin{eqnarray}
\hskip-0.5cm {d\pomega_1\over dt}\Bigr|_{\rm tide} & = & 
7.5 n_1 k_{2}\! {{1+\frac{3}{2}e_1^2 +
\frac{1}{8}e_1^4}\over{(1-e_1^2)^5}} {{M_*}\over{m_1}}\!\!
\left({{R_1}\over{a_1}}\right)^5,
\label{eq:pretide}\\ 
\hskip-0.7cm {d\pomega_1\over dt}\Bigr|_{J_2} \,\,& = & 0.5
n_1 {k_{2}\over{(1-e_1^2)^2}}\!\left({{\Omega_1}\over{n_1}}\right)^2\!\!
{{M_*}\over{m_1}}\!\!  \left({{R_1}\over{a_1}}\right)^5,
\label{eq:preJ2}\\
\hskip-0.5cm {d\pomega_1\over dt}\Bigr|_{\rm GR} & = & 3
n_1\, {{G M_*}\over{a_1 c^2 (1-e_1^2)}}.\label{eq:preGR}
\end{eqnarray}
\end{mathletters}
\noindent Note that in equation \refnew{eq:preJ2} the planet's spin rate,
$\Omega_1$, is scaled by its mean motion, $n_1$, since we expect that
tidal dissipation has produced approximate spin-orbit synchronization.
We neglect smaller precessional contributions due to the rotational
oblateness of the star and to the tidal distortion of the star by the
planet. 

Numerically, the total rate of extra precessions amounts to,
\begin{eqnarray}
\Delta & \equiv &
{d\pomega_1\over dt}\Bigr|_{\rm tide} +
{d\pomega_1\over dt}\Bigr|_{J_2} +
{d\pomega_1\over dt}\Bigr|_{\rm GR} \nonumber \\
& = & (7.79\, C \sin i + 1.52) \times 10^{-11}\s^{-1},
\label{eq:deltavalue}
\end{eqnarray}
where we set $\Omega_1 = n_1$ and introduce the dimensionless number
$C \equiv (k_{2}/k_{2J})(R_1/R_J)^5$, with $R_J$ and $k_{2J}$ assigned
values appropriate for Jupiter. Observations determine $k_{2J}=0.494$,
close to that of an $n=1$ polytrope, $k_2\approx 0.51$, whereas Saturn
has a smaller Love number, $k_2 = 0.317$ (Yoder \cite{yoder}) because
it is more centrally condensed as a result of possessing a relatively
larger core of heavy elements.

We incorporate the additional precessional terms into the secular
perturbation equations (see eq. [\ref{eq:secular}]). In a state of
apse alignment, there is no secular exchange of angular momentum
between the planets. Therefore, their orbital eccentricities are
constant on the secular time-scale. Moreover, the eccentricity ratio,
$e_1/e_2$, is a monotonically decreasing function of $\Delta$. To
match the observationally determined value of $e_1/e_2=0.19$ requires
$\Delta \approx 9.1\times 10^{-11}\,\sin^{-1} i\, \s^{-1}$. This value
is greater than the precession rate due to general relativity and
provides evidence for the tidal and rotational distortion of the inner
planet. Because the inner planet is not observed to transit, $\sin i <
0.99$, which implies $C \ge 0.9$. Nearly pole-on orbits, $\sin i <
0.20$, are excluded if we accept that $R_1 < 2 R_J$, a conservative
upper-limit as judged from Fig. 1 of Burrows \etal
\cite{burrows}. This implies $m_1 < 5.45 M_{\rm sat}$. 
Figure \ref{fig:secular_search} shows $e_1$ as a function of $C$ for
different values of $\sin i$ with $e_2$ fixed at its observed
value.

\vbox{
\begin{center}
\leavevmode \hbox{ \epsfxsize=0.85\hsize 
\epsffile[80 180 570 700]
{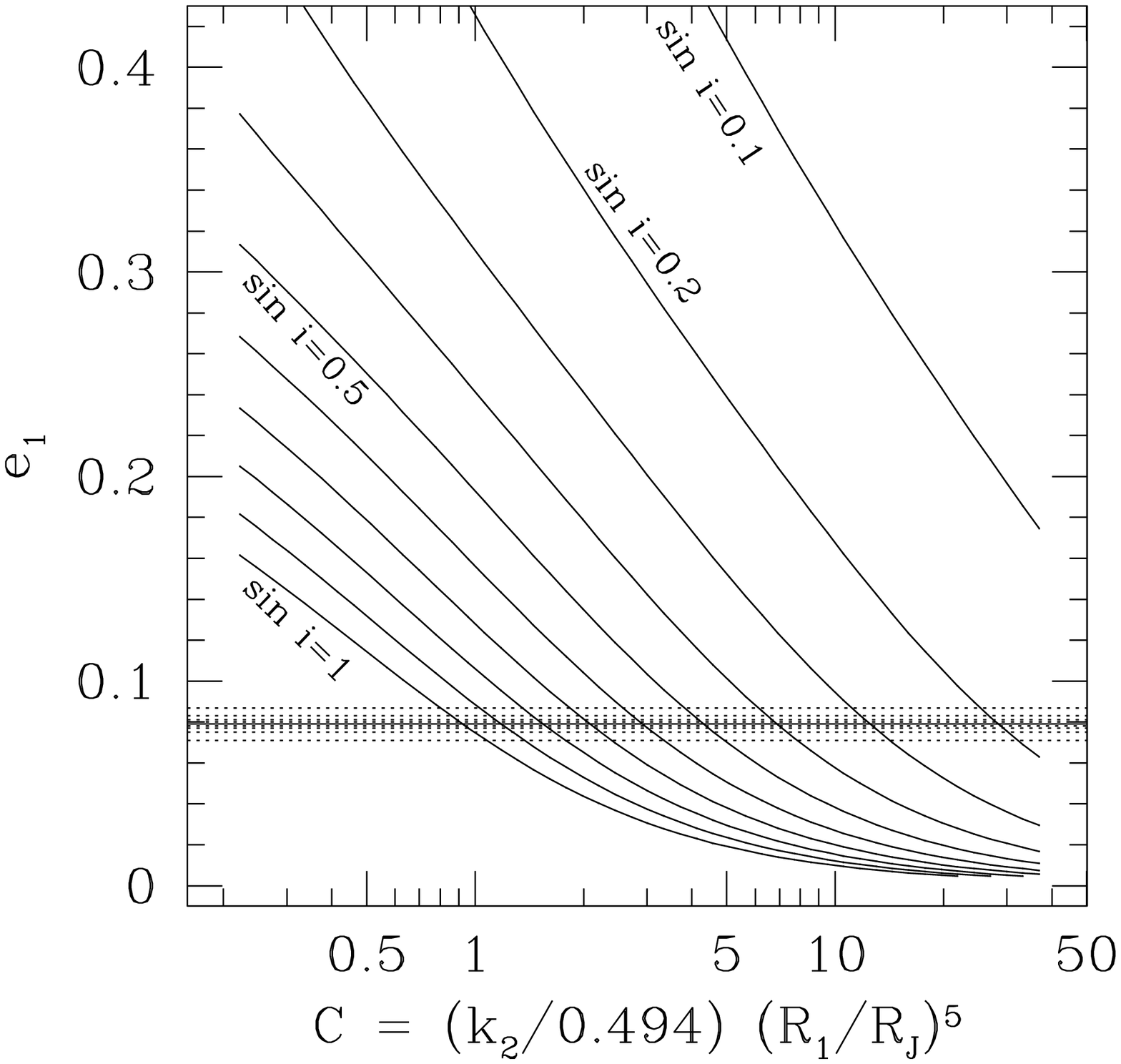}
} \figcaption{Values of the inner planet eccentricity, $e_1$,
as a function of $C$ for different values of $\sin i$.  The ten solid
lines are for $\sin i$ varying in equal increments from $0.1$ to
$1.0$. The horizontal band is centered on the measured value of $e_1$
and its vertical width spans a $1-\sigma$ error bar in each direction.
Relevant parameters are taken from Table
\ref{table:parameter}.\label{fig:secular_search}}
\end{center}}

\subsection{Tidal Dynamics}
\label{subsec:dampnonlinear}

Tidal dissipation in the approximately synchronously rotating inner
planet decreases its orbital energy without significantly changing its
orbital angular momentum. It follows from equations
\refnew{eq:dampingtau} and \refnew{eq:eJE} that this causes $a_1$ to
vary as $d\ln a_1/dt \approx - 2 e_1^2/\tau$. Since secular
interactions transfer angular momentum but not energy between the
planets' orbits, $a_2$ remains constant. Conservation of total angular
momentum leads to,
\begin{eqnarray}
{{dJ_1}\over{dt}} & = & J_1 \left({1\over 2} {{d \ln a_1}\over{dt}} -
{{e_1^2}\over{1-e_1^2}} {{d\ln e_1}\over{dt}}\right) \nonumber \\ & =
& - {{d J_2}\over{dt}} = J_2\,{{e_2^2}\over{1-e_2^2}} {{d\ln
e_2}\over{dt}},
\label{eq:Lconserv}
\end{eqnarray}
where $J_i = \sqrt{1-e_i^2}
m_i n_i a_i^2$. Apse alignment implies $e_1 = f(a_1, e_2)$, from which 
we deduce that
\begin{eqnarray}
{{d\ln e_1}\over{dt}}& =&  -{1\over\tau}\, 
{{f_{e_2} (1-e_1^2) + 2 e_2^2 f_{a_1} {{J_2 (1-e_1^2)}\over{J_1
(1-e_2^2)}}}\over{f_{e_2} +
{{J_2}\over{J_1}}\,
({{1-e_1^2}\over{1-e_2^2}})\,
({{e_2}\over{e_1}})^2}}, \nonumber \\
{{d\ln e_2}\over{dt}} & = &
- {1\over \tau}\, 
{{(1 - e_1^2) - 2e_1^2\, f_{a_1}}\over{f_{e_2} +
{{J_2}\over{J_1}}\,
({{1-e_1^2}\over{1-e_2^2}})\,
({{e_2}\over{e_1}})^2}},
\label{eq:relatea1e2}
\end{eqnarray}
where $f_{a_1} \equiv \partial\ln f/\partial\ln a_1$ and $f_{e_2}
\equiv \partial\ln f/\partial \ln e_2$. Numerically, $f_{a_1}
\approx 8$ and $f_{e_2} \approx 1$, so currently $e_1$ and $e_2$ are damping
on time-scales of $8.6\tau$ and $40\tau$, respectively; $e_1$ damps
faster than $e_2$ because the inner planet's semi-major axis is
shrinking.

\section{Linear Approximation}
\label{sec:linear}

In \S \ref{sec:nonlinear} we assumed that tidal dissipation had
delivered the planets' orbits into a state of periapse alignment. Here
we justify this assumption by studying the secular dynamics and 
tidal evolution under the somewhat inaccurate approximation of small
orbital eccentricities. 

Our starting point is the expansion of the secular disturbing function
to second order in $e_1$ and $e_2$. Substituting the expanded
disturbing function into Lagrange's equations for the variations of
the $e$'s and the $\pomega$'s, and adding a term to represent tidal
damping of $e_1$, yields the following set of linear equations
(MD2000)
\begin{equation}
{{dI_1}\over{dt}} = i A_{11} I_1 + i A_{12} I_2 - I_1/\tau,\hskip0.5cm
{{dI_2}\over{dt}} = i A_{21} I_1 + i A_{22} I_2,
\label{eq:Iincludetau}
\end{equation}
for the complex variable $I_i \equiv e_i \exp(i\pomega_i)$. The
coefficients $A_{ij}$ read:
\begin{eqnarray}
A_{11} & = & {{\alpha^2}\over{4}} {{m_2}\over{M_*}} n_1 B_1 + \Delta, 
\hskip0.2cm 
A_{12}  =  - {{\alpha^2}\over{4}} {{m_2}\over{M_*}} n_1 B_2,\nonumber \\
A_{21} & = & - {{\alpha }\over{4}} {{m_1}\over{M_*}} n_2 B_2, \hskip0.7cm 
A_{22}  =   {{\alpha }\over{4}} {{m_1}\over{M_*}} n_2 B_1,
\label{eq:defineAjk}
\end{eqnarray}
with $\alpha = a_1/a_2$, $B_1 = b_{3/2}^{(1)}$, and $B_2 = b_{3/2}^{(2)}$;
$b_s^{(j)}$ is the usual Laplace coefficient.
The symbols $\Delta$ and $\tau$ retain their definitions from \S
\ref{sec:nonlinear}.

In the absence of tidal damping ($\tau\to \infty$), the general
solution of equation \refnew{eq:Iincludetau} can be expressed as a
super-position of two eigenvectors ${\cal I}_*$ with
\begin{equation}
{\cal I}_* = \left[e_{1*}\exp(i\,\phi_{1*}), e_{2*}
\exp(i\,\phi_{2*})\right]\, \exp(i\,g_* t ),
\label{eq:defI}
\end{equation}
where
\begin{mathletters}
\begin{eqnarray}
\hskip-0.6cm g_*& = & {{(A_{11}+A_{22}) \pm \sqrt{(A_{11}-A_{22})^2 +
4 A_{12} A_{21}}}\over 2}\!\!, \\\hskip-0.6cm
{{e_{1*}}\over{e_{2*}}}\!\!&\! = &\!  \left|{{g_* -
A_{22}}\over{A_{21}}}\right| =\left|{A_{12}\over {A_{11}-g_*}}\right|,
\\\hskip-0.6cm \phi_{1*}\!\! &\! = &\!  \phi_{2*} +
{\pi\over 2}\left[1 + {\mbox{sgn}}(g_*-A_{22})\right].
\end{eqnarray} 
\label{eq:eigenvectors}
\end{mathletters}
We distinguish the two eigensolutions by $*=p$ and $*=m$ according to
the plus or minus sign that appears before the square root in the
expression for $g_*$.  Since $(e_{1p}/e_{2p})\, (e_{1m}/e_{2m}) =
A_{12}/A_{21} \approx 1.0$, one solution satisfies $e_{1} < e_{2} $
and the other $e_{1} > e_{2}$. If we take $\Delta = 5\times
10^{-11}\s^{-1}$, the $*=m$ solution reproduces the observed
eccentricity ratio $e_{1m}/e_{2m} = 0.19$ and apse
alignment.\footnote{This value of $\Delta$ is different from that
required in \S \ref{subsec:currenteratio} because the linear
approximation is inaccurate for $e_2 = 0.42$.}  The other solution has
$e_{1p}/e_{2p} = 5.26$ and anti-aligned apses. If both eigenvectors
have non-zero lengths, the orbital eccentricities oscillate out of
phase on time-scales of thousands of years ($\sim 1/g_*$) as depicted
in Mayor \etal (\cite{mayor83443}).

When $\tau$ is finite, $a_1$ and therefore $A_{ij}$ vary with
time. Since $\tau \gg 1/g_*$, the solution for the eigenvectors given by 
equation \refnew{eq:eigenvectors} remains valid. However, the eigenvector 
components, $e_{i*}$, decay slowly with time as
\begin{eqnarray}
\gamma_{1*} & = & {1\over \tau}\, {{1+2 e_{2*}^2\, f_{a_1}
 {{J_2}\over{J_1}}} \over{1+{{J_2}\over{J_1}}\,({{e_{2*}}
 \over{e_{1*}}})^2}},\nonumber \\ \gamma_{2*} & = &
 {1\over \tau}\, {{1-2 e_{1*}^2\, f_{a_1}}
 \over{1+{{J_2}\over{J_1}}\,({{e_{2*}}
 \over{e_{1*}}})^2}},
\label{eq:ggamma}
\end{eqnarray}
where $\gamma_{i*} \equiv - d\ln e_{i*}/dt$.  In the limit of small
eccentricity, the expressions for the nonlinear damping rates given by
equation \refnew{eq:relatea1e2} reduce to those in equation
\refnew{eq:ggamma} provided $f_{e_2}=1$ as is appropriate for the
linear problem. Each of the four eigenvector components has its
individual damping rate. Both components of the $p$ eigenvector decay
more rapidly than those of the $m$ eigenvector. If the current state
consists mostly the $m$ eigenvector with a small admixture of the $p$
eigenvector, both components in the latter would decay on time-scale
$\approx \tau$, while the dominant components $e_{1m}$ and $e_{2m}$
would decay much more slowly, as $\gamma^{-1}_{1m}\approx 7\tau$ and
$\gamma^{-1}_{2m}\approx 32\tau$.

It is notable that the asymmetry in damping rates between the aligned
and the anti-aligned solutions only applies if the inner planet has an
orbital period $\leq 5$ days. For longer periods, the additional
precession rates (eqs. [\ref{eq:pretide}] - [\ref{eq:preGR}]) are
unimportant and $(e_{1m}/e_{2m}) \sim (e_{1p}/e_{2p}) \sim 1$.

\section{Conclusions}
\label{sec:conclusion}

Both the survival of HD $83443b$'s orbital eccentricity and the
locking of its apse to that of HD $83443c$ are straightforward
consequences of secular and tidal dynamics. The ratio $e_1/e_2$
provides compelling evidence for the tidal and rotational distortion
of HD $83443b$. It constrains unknown system parameters by relating
$C\equiv (k_{2}/k_{2J})(R_1/R_J)^5$ to $\sin i$.  In particular, we
find $C > 0.9$.

Apse alignment indicates that a minimum amount of tidal dissipation
has occurred over the life-time of the system, taken to be $N$ Gyrs.
By requiring that this is at least 3 times longer than $\tau$, we
deduce that $Q < 5 \times 10^5\, N C \sin i$.  It may also be possible
to infer a lower limit for $Q$ by integrating the orbits backward in
time.

At the present time $n_1/n_2 = 9.99 \pm 0.06$, suggestive of a $10:1$
mean motion resonance. This is probably just a coincidence. Moreover,
no physically important effects would be associated with this
resonance should it exist.

\begin{acknowledgements}
This research was initiated while Y. Wu participated in a CITA-ITP
postdoctoral exchange program funded by NSF Grant PHY-9907949. She
thanks Drs. Man-Hoi Lee, Stan Peale, and Brad Hansen for helpful
discussions. P. Goldreich's research is supported by NSF Grant
AST-0098301.
\end{acknowledgements}

{}

\begin{appendix}

\section{Secular Perturbations}
\label{sec:secular}

Secular changes of the inner planet's periapse angle, $\pomega_1$, and
eccentricity, $e_1$, are described by
\begin{equation}
{de_1\over dt}  =  {de_1\over dt}\Bigr|_2, \hskip1.5cm
{d \pomega_1\over dt}  =   {d \pomega_1\over dt}\Bigr|_2 + {d\pomega_1\over 
dt}\Bigr|_{\rm tide} +
{d\pomega_1\over dt}\Bigr|_{J_2} +
{d\pomega_1\over dt}\Bigr|_{\rm GR}, 
\label{eq:secular}
\end{equation}
where the subscript $2$ denotes effects arising from secular
interactions with the outer planet.
Similar equations apply to the evolution of $\pomega_2$ and $e_2$ but
the only significant perturbations are those from the inner
planet. Expressions for the precession rates due to the tidal and
rotational distortion of the inner planet and to general relativity
are given in the main body of the text. We obtain $d\pomega_1/dt|_2$
by applying a method devised by Gauss (\cite{gauss}) as outlined
below.

Consider a pair of coplanar rings formed by spreading the mass of each
planet along its orbit such that the line density is inversely
proportional to the orbital velocity. We decompose into radial, $\cal
R$, and tangential, $\cal T$, components the gravitational force
exerted on a line element of length $r_1 df_1$ by a line element of
length $r_2 df_2$, where $f$ denotes true anomaly and $r$ distance
from the star; 
\begin{equation} 
{\cal R} = {{G \rho_1 \rho_2 r_1 r_2
df_1 df_2}\over {|\boldr_1 - \boldr_2|^2}} \cos\theta, \hskip1.5cm
{\cal T} = {{G \rho_1 \rho_2 r_1 r_2 df_1 df_2}\over {|\boldr_1 -
\boldr_2|^2}}\sin \theta,
\label{eq:radialtan}
\end{equation}
with $\theta$ the angle between $\boldr_1$ and $\boldr_2 -
\boldr_1$. We remove $\theta$ in favor of $\phi$, the angle between
$\boldr_1$ and $\boldr_2$ ($\phi = f_2 + \pomega_2 - f_1 -
\pomega_1$), by applying the geometric relations,
$|\boldr_1-\boldr_2|\, \sin \theta = r_2 \sin \phi$ and $|\boldr_1 -
\boldr_2|\, \cos\theta = r_2 \cos \phi - r_1$. Then we integrate over
$df_2$ to obtain the components of the total force due to planet 2 on
the mass of planet 1 in the line element $r_1 df_1$. When this result
is substituted into Gauss' form of the perturbation equations (Gauss
\cite{gauss}) and integrated over $f_1$, we obtain\footnote{Our
derivation is slightly different from that given in MD2000, but the
results are identical.}
\begin{eqnarray}
{d \pomega_1\over dt}\Bigr|_2 & = & {{\sqrt{1-e_1^2}}\over{m_1 n_1 a_1
e_1}}\, \oint df_1 \, \left[ - \cos f_1 \oint df_2\, {\cal R} + {{
(2+e_1 \cos f_1)\, \sin f_1}\over{1+e_1 \cos f_1}} \oint df_2 \, {\cal
T} \right], \nonumber \\
{de_1\over dt}\Bigr|_2 & = & {{\sqrt{1-e_1^2}}\over{m_1
n_1 a_1}}\,\oint df_1\, \left[ \sin f_1 \oint df_2 \,{\cal R} +
{{(2\cos f_1 + e_1 +e_1 \cos^2 f_1)}\over{1+e_1 \cos f_1}} \oint
df_2\, {\cal T}\right].
\label{eq:gauss}
\end{eqnarray}
We have verified that when integrating these equations with parameters
appropriate to HD $83443$'s planetary system, we obtain results that
agree well with those obtained by direct integration of the orbits
using the SWIFT package (Levinson \& Duncan \cite{swift}).

\end{appendix}

\end{document}